# 20 GHZ LOW NOISE LLRF SYSTEM

A. Rohlev (TSR Engineering)

*Abstract*

A 20 GHz LLRF system is being built using a two-board (RF Front End + ADC/DAC/FPGA) architecture. The RF Front End provides 8 down-converting channels and 3 up-converting channels (5.5–20 GHz RF ↔ 0.05–3 GHz IF). Separate, phase locked, low-noise input and output LO's are generated on-board with an independent programmable frequency range of 4–20 GHz. A user input is provided so that both LO's as well as all ADC, DAC, and FPGA clocks can be locked to a supplied reference source with a frequency range from 100 MHz to 20 GHz. The IF is processed with a commercial board (HiTech Global ZRF8) based on the Xilinx ZYNQ RFSoC FPGA. The RFSoC FPGA incorporates eight 4-GSPS 12-bit ADC's with a 4 GHz analog bandwidth and eight 6.4-GSPS 14-bit DAC's. The ZRF8 is a PCIe-standard board that provides low noise ADC/DAC/FPGA clocking, 16 GB of memory, a FMC+ socket, and a 1 Gbps Ethernet port. The complete system will be housed in a standard 2U 19" rack.

## INTRODUCTION

Two significant forces have been pushing LLRF development in recent years. First is the quest for higher frequency accelerators. Whether from the demands of physics (in the case of FELs) or simply the physical demands (shorter and therefore less expensive accelerators), cavity frequencies have steadily been climbing from S band (3 GHz) to C band (6 GHz) to X band (12 GHz) and beyond. Consequently both cavity control as well as monitoring systems (BPMs, Wakefield) have been moving to higher input and output frequencies. This has driven the development of higher and higher frequency RF front ends.

The second development is the introduction of multiple ultra-high speed ADCs and DACs integrated directly with a very high performance FPGA – wherein the data enters the fabric directly eliminating the need for multiple chips with external timing and complex high speed serial routing. This leads to higher performance, lower power, lower complexity, and potentially lower cost processing boards.

We aim to combine both of these developments into a single LLRF/monitoring system with a 20 GHz front-end and an ultra-high speed integrated ADC/DAC/FPGA processing board.

## ARCHITECTURE AND HARDWARE

As with other high frequency LLRF designs [1] the 20 GHz system utilizes a two-board architecture: A RF Front End (RFFE) coupled to a ADC/DAC FPGA Processing Board (PB) both housed in a 19" 'pizza box' chassis as shown in Figure 1. This approach provides the highest signal integrity, often at the lowest price, at the expense of space and standardization. The 20 GHz LLRF system couples a custom RFFE to a commercial processing board [2] built around the new RF system-on-chip (RFSoC) FPGA series from Xilinx.

The RF front end has 8 down-converting input channels and 3 up-converting output channels. Independent low noise, PLL-based, LO generation and distribution is provided for the input and output channels. The input and output LO's are phase locked to the same external 100 MHz OCXO which can itself be locked to a user provided reference in the range of 100 MHz to 20 GHz.

Bandwidth limitation of the amplifiers and power splitters constrain both the input and output LO generation to a frequency range of 5.5 to 18 GHz. The IF for both input and output channels is AC coupled with a lower limit of 50 MHz. Thus the effective system bandwidth, for both inputs and outputs, is 5.5 – 20 GHz for an IF to/from the processing board of 50 MHz to 2 GHz.

The RFFE incorporates numerous diagnostics for LO power levels, RF output levels, DC voltages and currents as well as an array of temperature sensors. A microcontroller interfaces to all LO generation and reference PLL's as well as providing a USB interface to the controller board. The RFFE also houses 6 TEC controllers with a combined 40W Peltier heating/cooling capacity that can be used to stabilize the board temperature within ±0.5°C. The RFFE is specifically designed for low noise and low leakage. Its RF sections are fully shielded and insulated. All amplifier, mixer and PLL DC power is provided thru ultra-low noise linear regulators themselves buffered by LDO's from the on-board switch-mode convertors. Meanwhile all high-noise functions, like the switch-mode convertors and TEC controllers, are hosed on separate, individually shielded daughter cards. All RF and IF connectors are SMA.

The Processing Board is the HTG-ZRF8 from HighTech Global. It is a commercially available PCIe standard card built around the XCZU28DR RFSoC FPGA from Xilinx. The RFSoC FPGA incorporates eight 12-bit 4-GSPS ADC's with 4 GHz analog bandwidth as well as eight 14-bit 6.4 GSPS DAC's. All IF signals are on SSMC connectors. Furthermore the ZRF8 hosts up to 16GB of DDR4 memory, one FMC+ (Vita 57.4) socket, and various data interfaces - SATA, Display Port, USB3, UART, as well as a 10/100/1000 Mbps Ethernet port. The ZRF8 utilizes a low-noise fully programmable PLL synthesizer (TI's LMX2582) to drive the FPGA's ADC/DAC clocking port. This PLL is referenced/locked to the same 100 MHz OCXO as the PLL's on the RFFE thus ensuring synchronization between the IF signals and the ADC/DAC clocks. The FPGA's core clocks, serial data clocks, DDR clocks, and FMC+ clocks are generated from another programmable synthesizer itself locked to an output of the LMX2582.

A calibration board has been developed for the input channels. It allows a known signal to be injected, in situ, into all eight input channels. The calibration signal can be generated on board or externally. The amplitude of the calibration signal can be kept constant over a wide frequency range thus allowing for the measurement and calibration of variations in the input chain (attenuator / filter / mixer / transformer / filter / amplifier / cable / transformer / ADC) with temperature and time. The calibration values can be stored in a look-up table in the FPGA and used to flatten the frequency response of each channel. Furthermore, if Output 2 of the RFFE is used as the calibration source, $3^{rd}$ order harmonic distortion (OIP3) can be measured and calibrated since this output combines two signals (after up-conversion and amplification) allowing for a two-tone excitation. On the down side the presence of the calibration board introduces a path by which leakage from adjacent channels can contaminate the signal. The channel-to-channel isolation, with the calibration board present, is approximately 63 dB.

The entire system (RFFE, ZRF8, Calibration board, power supplies, external OCXO) can fit in a 2U or 3U 19" enclosure with a standard depth of 43.5 cm. Power to both the RFFE and ZRF8 is supplied from a single 13A 12V switch-mode supply. Furthermore a ±15V 0.5A linear supply is needed for the external OCXO and RFFE amplifier bias voltages. The RFFE is connected to the ZRF8 thru standard coax cables. Semi-flex cables are used for the input reference and calibration drive signals. The RFFE can drive up to 10 bi-colored LEDs thru a standard socket as indicators. Front and rear panel 12V fans provide airflow and all external communication is thru a RJ45 connector to the ZRF8's Ethernet port.

## LOW NOISE LO GENERATION AND CLOCKING

To achieve optimal ADC noise performance (as in its datasheet SNAD values) the jitter on the acquisition clock, and LO's if down-conversion is used, must be considerably less than the ADC's internal jitter level at a given input frequency which can be calculated from the equation below [3].

$$t_a = \frac{1}{2\pi f \cdot \text{ALOG}\left(\frac{\text{SNAD}}{20}\right)}$$

SNAD (dBFS) is the signal to noise and distortion value from the ADC datasheet, at the relevant input frequency, $f$ (Hz). The resulting equivalent convertor jitter is $t_a$ (s).

The ADCs in the Xilinx FPGA are 12-bit devices that sample at 4 GSPS with an analog bandwidth of 4 GHz. The datasheet [4] lists that noise spectral density (NSD) at a set input frequency and sampling rate from which the SNAD value can be calculated. For an input frequency of 2.4 GHz and a sampling rate of 4 GSPS the NSD value is -150 dBFS/Hz which corresponds to a SNAD value of 57 dB (ENOB = 9.2). Using the equation above this corresponds to an equivalent ADC jitter of 94 fs. The clock/LO jitter is then added to this value, in quadrature, to calculate the final jitter and subsequent degradation of SNAD/ENOB.

Both input and output LO's are generated from a parallel 20 GHz synthesizers (TI LMX2595) which are locked to the same external 100 MHz OCXO (Wenzel Citrine Gold). Simulation show that with a passive $4^{th}$ order loop filter an integrated phase jitter of 56 fs [10 Hz – 100 MHz] can be achieved for a 15 GHz output from each synthesizer. By summing the outputs of two identically configured synthesizers [5] the phase noise is reduced by 3 dB across the spectrum resulting in a jitter value of approximately 40 fs.

To calculate the effect of 40 fs LO jitter on the ADC performance we add it, in quadrature, to the internal ADC jitter of 94 fs to get a sum of 102 fs. Using this value as $t_a$ in Equation 1. results in a SNAD of 56.25 dBFS or an ENOB of 9.05 bits – a degradation of only 0.15 bits.

To achieve this relatively low phase jitter at 15 GHz it is critical that the 100 MHz reference oscillator have extraordinarily low phase, especially at offsets below 1 kHz. The Citrine Gold model from Wenzel achieves -138 dBc/Hz at 100 Hz offset. It also serves as the reference oscillator for the ADC/DAC clock synthesizer on the ZRF8 board thus ensuring the lowest possible phase noise on the ADC clocks.

## HIGH SPEED VS. HIGH ACCURACY ACQUISITION

The ADCs on the Xilinx RFoC FPGA are very fast (4GSPS), with a very high analog bandwidth (4 GHz), but they are not very accurate (ENOB = 9.2 @ 4 GSPS, 2400 MHz input frequency). However, their speed can be exchanged for accuracy by oversampling the input signal. To illustrate this we will compare the accuracy of the RFSoC ADCs against the LTC2107, the most accurate high speed ADC on the market (ENOB = 12.8 @ 210 MSPS, 140 MHz input frequency), at input frequencies of 100 and 500 MHz.

For a 100 MHz input signal the LTC2107 sampling at 210 MSPS has a SNAD of 78.7 dBFS corresponding to 12.8 effective bits or an intrinsic jitter of 185 fs. For the same 100 MHz input signal the RFSoC ADC sampling at 4 GSPS has a SNAD of 60 dBFS corresponding to 9.7 effective bits or an intrinsic jitter of 1592 fs. However, if the extra samples of the RFSoC are averaged the intrinsic jitter can be reduced by factor of SQRT(4000/210) = 4.36 leading to a value of 364 fs. This corresponds to a SNAD of 72.8 dBFS or 11.8 effective bits – exactly one bit less than the LTC2107.

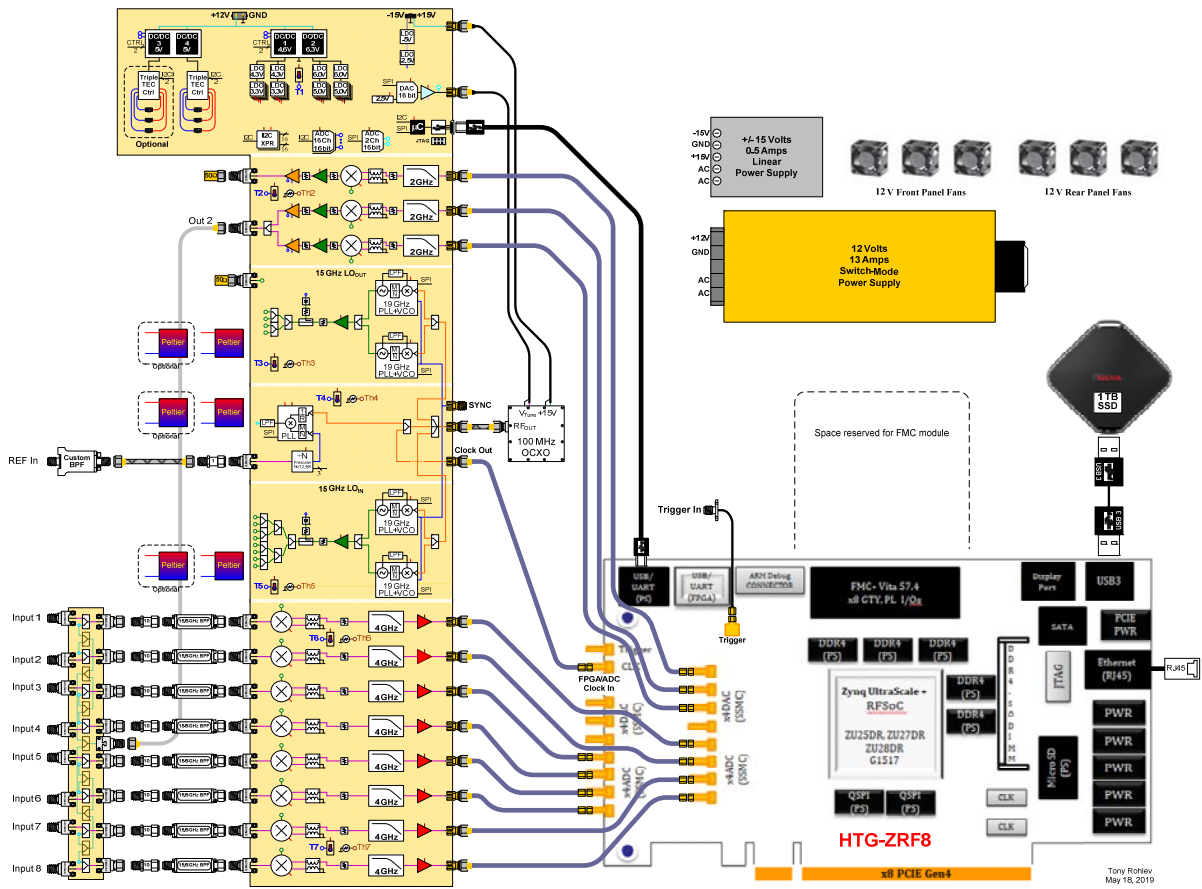

Figure 1: Block diagram of 20 GHz LLRF system.

At 500 MHz input the LTC2107 has a SNAD of 75.1 dBFS which corresponds to 12.2 effective bits or an intrinsic jitter of 56 fs. The RFSOC ADC at 500 MHz input still has a SNAD of 60 but at the higher frequency this now corresponds to an intrinsic jitter of 318 fs. Dividing this by the same factor of the SQRT(4000/210) leads to an intrinsic jitter of 73 fs corresponding to a SNAD of 72.8 dBFS or the same 11.8 effective bits. Thus at 500 MHz the difference between the most accurate high speed ADC on the market and the RFSoC is less than half a bit. The analog bandwidth of the LTC2107 is limited to 700 MHz thus comparisons at 1 GHz or more, where the RFSoC will be operating, can't be done.

As a final note we can compare the performance of ADCs on the Fermi LLRF System [1] (which has demonstrated consistent control of 3 GHz pulsed RF in normal conducting cavities to within 0.1°, 0.1%) with the RFSoC ADCs. At Fermi the LTC2209 is used with an input frequency of 99 MHz and a sampling rate of 120.8 MSPS. At 99 MHz the LTC2209 has a SNAD of 75.7 dBFS corresponding to a ENOB of 12.8. The RFSoC still has a SNAD of 60 dBFS but due to the higher sampling ratio (4000/120.8) the effective number of bits are now 12.7, only a tenth of a bit below the Fermi system.

## CONCLUSION

The hardware system described above is currently under development and we look forward to providing test results in the near future. Future systems connecting to very high channel count processing boards like HTG's ZRF16 [] are being considered.